\documentclass{aa}
\usepackage{graphicx}

\begin{document}

\title{Compact strange stars with a medium dependence in gluons at finite
temperature}
\titlerunning{Strange compact star ..}
\author{ Manjari Bagchi \inst{1} \and Subharthi Ray \inst{2}
\and Mira Dey \inst{1,3} \and Jishnu Dey \inst{1}}
%\footnote{email : kamal1@vsnl.com.}.}
\authorrunning{Bagchi et al.}

\offprints{M. Bagchi}

\institute{Dept. of Physics, Presidency College, 86/1, College
Street, Kolkata 700073, India \\ \and Inter University Centre for
Astronomy and Astrophysics (IUCAA), Post Bag~\#~4, Ganeshkhind,
Pune 411007, India \\ \and CSIR Emeritus Professor }

\date{}

\abstract {The possible existence of  strange stars in the
universe will help in the understanding of various properties of
quantum chromodynamics, like asymptotic freedom and chiral
symmetry restoration, which is otherwise very difficult to prove
in laboratory experiments. Strange star properties were calculated
using large color approximation with built-in chiral symmetry
restoration. A relativistic Hartree Fock calculation was
performed using the Richardson potential as an interquark
interaction. This potential has the asymptotic freedom and a
confinement-deconfinement mechanism built into it and the present
calculation employs an application of this potential with modified
two scale parameters $\Lambda$ and $\Lambda^{\prime}$, to find a
new set of equations of state for strange quark matter. The linear
confinement string tension from lattice calculations is 350 $MeV$
and the Coulomb -like part has the parameter 100 $MeV$ from deep
inelastic scattering experiments. We also consider the effect of
temperature, on gluon mass in a simple way, in addition to the
usual density dependence. We obtained a set of new equation of
states for strange quark matter and also found that the transition
temperature from hadronic matter to strange matter is at 80 MeV,
close to the 100 MeV estimated in litarature.Formation of strange
stars may be the only signal for formation of quark-gluon plasma
with asymptotic freedom and chiral symmetry restoration and this
may be observable through many processes -such as for example
through delayed $\gamma$ ray afterglow.}

\maketitle

\section{Introduction}

Two decades have passed since the existence of strange matter and
strange stars was proposed by Witten (1984), and it is still
difficult to prove or disprove the existence of a state of strange
quark matter in its purest form. Besides the mass radius relation,
efforts have been made to find other signatures for the existence
of quark stars. The existence of absorption bands at 0.35 $keV$
with harmonics at 0.7, 1.05, 1.4 etc, in 1E 1207$-$5209 was
interpreted in \cite{mpla} as evidence for strange stars. It was
then observed that six stars show harmonic band emission at
energies corresponding to the same compressional modes at which
the absorption bands were invoked (\cite {MNRAS}). The SPI
spectrometer on the INTEGRAL satellite detected a bright 511 keV
line in the $\gamma$ - ray spectrum from the bulge of the galaxy
with a spherically symmetric distribution (\cite{knodjean}). This
has stimulated research to the fundamental physics that describes
cosmological dark matter. Several authors (\cite{bhscp04}, Oaknin
and Zhitnitsky, 2005) have suggested that the line comes from
positron annihilation, possibly in strange stars made of
antiquarks. The implication of cosmic separation of phases
(Witten, 1984) is that strange stars (SS) can exist since the
early universe and can themselves explain dark matter. A crucial
condition for the survival of these stars, given by Alcock and
Olinto (1989), is a large value of surface tension
$\sim(178~MeV)^3$ which can be obtained from strange star equation
of state (EOS) (\cite{oursurf}). In his recent review, Weber
(2005) argued that the constraints on mass radius relations from
observational data support that the compact objects SAX
J1808.4$-$3658, 4U 1728$-$34, RX J1856.5$-$3754 and Her X$-$1 are
strange stars rather than neutron stars and on the basis of its
low value of temperature, PSR 0205$+$6449 may be considered as a
strange star. He also discussed the possibility of explaining some
astrophysical phenomena like X-ray bursts, gamma-ray bursts and
soft gamma repeaters with a strange star model. The most
controversial is the isolated compact star RX J1856.5$-$3754,
which has a featureless thermal spectrum, and its X-ray data fits
quite well with a blackbody distribution. With a distance estimate
of 120 pc, Drake et al. (2002) gave the radiation radius of the
star to be 4.12$\pm$0.68 km. This indicates that the stellar
radius is 4 km to 8 km. Later, with a better estimation of the
distance of 175 pc, the estimated radiation radius $R^\infty
\simeq 8$ km. However, there are controversies regarding the
surface temperature of this star. With an estimated surface
temperature ($T^\infty$) of 63.5 $eV$, $R^\infty \simeq 4.4$ km,
and with $T^{\infty}  \simeq 32 eV$, $R^\infty \simeq 8$ km
(\cite{bur03}).

There is evidence, as recently discussed by Bombaci, Parenti and
Vidana (2004) - for the existence  of compact strange and neutron
stars with larger radius and a shift from the latter to the former
can account for delayed gamma ray bursts.

PSR 1913+16 is the first  pulsar to be discovered in a double
neutron star system is, also known as the Hulse Taylor pulsar
(1975). It is estimated to have a mass of 1.441 M$_\odot$ and the
recently discovered pulsar (Lyne et al., 2004) PSR~J07373039A,
forming a binary pulsar system, has a mass of 1.337 M$_\odot$.
Thus the mass of pulsars may lie near these values, the observed
variation of periodicity of the pulsars indicates the variation
in their masses. Although it may be a stringent lower-limit
constraint for neutron stars masses, which are bound mainly by
gravitational forces, the same does not hold for compact objects,
which are self bound, like a system of quark matter.
Theoretically there is no physical reason to have a lower limit
for the mass of strange stars, where strong interactions play an
important role in stellar binding. The same EOS will allow even a
very small quark nugget.

In the literature, there are several EOSs for strange quark
matter, starting from the Bag model (\cite{afo86}, \cite{han86},
\cite{ket95}) to the recent models like the mean field model with
interacting quarks (\cite{D98}), the strange star EOS model in the
perturbative quantum chromodynamics (QCD) approach (\cite{fps01}),
those of the chiral chromodielectric model (\cite{mft03}) or
Dyson-Schwinger model (\cite{blas99}), etc. In our present work,
we will consider only the relativistic mean field model by Dey et
al. (1998).

In the relativistic Hartree-Fock calculation of a strange quark
matter EOS to form a compact star (\cite{D98}), a phenomenological
inter-quark interaction, namely the Richardson potential, was
used. Strange stars are more compact than conventional neutron
stars and fit into the Bodmer (1971) -- Witten (1984) hypothesis
for the existence of strange quark matter. They were able to
produce a very compact stellar structure with a mass of about 1.44
$M_\odot$ and radius of about 7 km. The original Richardson
(1979) potential was designed to obtain the mass spectrum of
heavy mesons (Charmonium and Upsilon). It includes two features of
the qq force, the asymptotic freedom (AF) and confinement,
however, with the same scale, ${\Lambda}$.

The ${\Lambda}$ parameter needed was 400 MeV. It was further
applied to light meson spectroscopy  and baryon properties with
the same value of ${\Lambda}$ (Crater and van Alstine, 1984, Dey,
Dey and Le Tourneux, 1986). Although both the confinement and AF
scales are chosen to be the same in the Richardson potential, it
is the confinement, not the AF part, that plays the more important
role in determining the static properties of hadrons. However,
when the AF part is important (as in strange star property
calculations), ${\Lambda}$ would be much smaller, around 100 MeV,
as it is the scale for AF, obtained from perturbative QCD. Indeed,
for self bound high density strange quark matter, Dey et al.
(1998) found that ${\Lambda}$ needed to be $\sim 100 $ MeV. In
the present paper, proper scales are chosen for both the
parameters, as determined from a calculation by Bagchi et al.
(2004) by fitting the masses and recently determined magnetic
moments of $\Omega^-$ and $\Delta^{++}$.

In section~\ref{sec:model} we describe the basic formalism of the
model as given in Dey et al. (1998) and the calculations
incorporating finite temperature effects. In
section~\ref{sec:modify} we present a fine tuning of the Dey et
al. (1998) model, with a modified version of the Richardson
potential, and also incorporate a temperature dependence of gluon
mass. In section~\ref{sec:result} we show the results and the
plots, and draw conclusions and summarize in
section~\ref{sec:conc}.

\section{The model}\label{sec:model}
There are no free parameters in QCD with which one can
systematically build up a perturbation series. So \cite{thooft}
suggested use of the inverse of the number of {\it{colors}} -
which is the key quantum number in QCD - as an expansion
parameter. In this case a baryonic system like stars can be
explored in relativistic tree level calculations with a potential
that may be obtained from the meson sector phenomenology
\cite{witten}. In this connection the Richardson potential is a
very good choice for interquark interactions.

 The original Richardson potential is given by:
\begin{equation}
V_{ij} = \frac{12 \pi}{27}\frac{1}{ln(1 +  {({\bf k}_i- {\bf
k}_j)}^2 /\Lambda ^2)}\frac{1}{{({\bf k}_i - {\bf k}_j)}^2}
\label{eq:V}
\end{equation}

At high density (high Fermi momentum) quarks are deconfined as
the bare potential is screened. Confinement is softened and the AF
part takes over. This bare potential in a medium will be screened
due to gluon propagation and  $({\bf k}_i - {\bf k}_j)^2$ of
equation (\ref{eq:V}) is replaced by $[{({\bf k}_i - {\bf
k}_j)^2}+D^{-2}]$. The inverse screening length, $D^{-1}$, which
can be identified with the gluon mass $m_g$, to the lowest order
is :
\begin{equation}
(D^{-1})^2 ~=~ \frac{2 \alpha_0}{\pi} \sum_{i=u,d,s,}k^f_i
\sqrt{(k^f_i)^2 + m_i^2} \label{eq:gm}
\end{equation} where $k^f_i$, the Fermi momentum of the {\it i-th} quark is
obtained from the corresponding number density:
\begin{equation}
{k^f_i} = (n _i \pi^2)^{1/3} \label{eq:kf}
\end{equation}
and $\alpha_0$ is the perturbative quark gluon coupling.

In this model, chiral symmetry restoration at high density is
incorporated by assuming that the quark masses are density
dependent :

\begin{equation}
M_i = m_i + M_Q sech\left( \frac{n_B}{N n _0}\right), \;\;~~~ i =
u, d, s \label{eq:qm}
\end{equation}
where $n_B = (n_u+n_d+n_s)/3$ is the baryon number density; $n_0
= 0.17~fm^{-3}$ is the normal nuclear matter density; $n_u$,
$n_d$, $n_s$ are the number densities of u, d and s quarks
respectively and $N$ is a parameter. The current quark masses
($m_i$) are taken as : $m_u = 4 \;MeV,\; m_d = 7 \;MeV,\; m_s =
150 \;MeV$. It is ensured that in strange matter, the chemical
potentials of the quarks satisfy ${\beta}$ equilibrium and charge
neutrality conditions. The parameters $M_Q$ and $N$ are adjusted
in such a way that the minimum value of $E/A$ for u,d,s quark
matter is less than that of Fe$^{56}$, so that u,d,s quark matter
can constitute stable stars. The minimum value of $E/A$ is
obtained at the star surface where the pressure is zero. The
surface is sharp since strong interaction dictates the
deconfinement point. An electron cloud forms above this sharp
surface - since they are not held by strong interaction. However,
the minimum value of $E/A$ for u-d quark matter is greater than
that of Fe$^{56}$ so that Fe$^{56}$ remains the most stable
element in the non-strange world. With the obtained EOSs, they
solved the hydrostatic equilibrium equations
(Tolman-Oppenheimer-Volkov equation) to get the structure of the
stars.

Finite temperature $T$ was incorporated in the system through the
Fermi function (Ray et al., 2000b):
\begin{equation}
FM(k,T)=\frac{1}{e^{(\epsilon - \epsilon_F)/T} + 1}
\end{equation}
with the flavour dependent single particle energy
\begin{equation}
\epsilon_i=\sqrt{k^2+M_i(\rho)^2}+U_i(k). \label{eq:ep}
\end{equation}

\begin{equation}
I=\frac{g}{2\pi^2}\int_0^\infty \phi(\epsilon)k^2FM(k,T) dk
\label{eq:i}
\end{equation}
where $g$ is the spin-colour degeneracy, equal to 6 and $I$ is the
number density for $\phi(\epsilon) = 1 $ and energy density for
$\phi(\epsilon) = \epsilon  $.

The system is highly degenerate even at very high $T$ which is
around 70 $MeV$ since the chemical potential is very high, of the
order of several hundred $MeV$. The entropy was calculated as
follows:
\begin{eqnarray}
\nonumber
S(T)&=&-\frac{3}{\pi^2}\int_0^\infty k^2[FM(k,T)ln(FM(k,T)) \\
&& + (1-FM(k,T))ln(1-FM(k,T))] dk \label{eq:s}
\end{eqnarray}
The pressure was calculated from the free energy
\begin{eqnarray}\nonumber
f = \epsilon - Ts
\end{eqnarray}
as follows:
\begin{equation}
P=\sum_i\rho_i\frac{\partial f_i}{\partial \rho_i} - f_i
\label{pressure}
\end{equation}

Ray et al. (2000b), found that stable star structure cannot be
obtained with a temperature greater than 70 $MeV$. But here the
effect of temperature on gluon mass was not considered. Also, the
parameter ${\Lambda}$ needs to be different. Can one separate the
confinement  and AF part and apply the same two parameters to
hadron properties as well as to high density quark matter?
Moreover, what happens to the strange star properties if the
effect of temperature on gluon mass is considered?

Successful attempts have been made to construct a two parameter
Richardson potential to explain energies and magnetic moments of
baryons. Recently the magnetic moments of  $\Delta^{++}$ and
$\Omega ^-$ with three $u$ and three $s$ valence quarks
respectively have been accurately found in experiments. These
sensitive properties are fitted with a modified Richardson
potential with different scales for confinement ($\sim 350~ MeV$)
and AF ($\sim 100 ~ MeV$) (Bagchi et al., 2004). It was also shown
that baryonic properties depend much on the confinement part and
less on the AF part as baryons are confined quark systems. It is
natural to apply this new potential to star calculations since
they are constrained by baryonic data. Retaining the approach of
Dey et al. (1998), we used this modified Richardson potential to
derive our EOS.

\section{Modifications}\label{sec:modify}

The properties of AF and confinement of quarks in the model is
embedded in the potential we have used, i.e., the Richardson
potential. We have extensively searched for the two parameter
form of this potential that would separately reflect the
properties of AF and quark confinement. In the light of the new
modifications, the Richardson potential takes the form:

\begin{eqnarray}
\nonumber V_{ij} &=& \frac{12 \pi}{27}\left[\frac{1}{{\rm ln}(1 +
\frac{{({\bf k}_i- {\bf k}_j)}^2}{\Lambda
^2})}-\frac{\Lambda^2}{({\bf k}_i- {\bf k}_j)^2}
+\frac{{\Lambda^\prime}^2}{({{\bf k}_i- {\bf k}_j})^2}\right]\\
&& \times \frac{1}{({\bf k}_i - {\bf k}_j)^2} \label{eq:Vtl}
\end{eqnarray}
with ${\Lambda}^{ \prime}$ the confinement property and
${\Lambda}$ asymptotic freedom.

The term $\left(\frac{1}{Q^2ln(1 +  Q^2 /\Lambda
^2)}-\frac{{\Lambda}^2}{Q^4}\right)$ is asymptotically zero for
large momentum transfer $Q^2 = ({{\bf k}_i- {\bf k}_j})^2$ and
the term $\frac{{{\Lambda}^{\prime}}^2}{Q^4}$ explains the
confinement reducing to a linear confinement for small $Q^2$.

Previously Bagchi et al. (2004) used the same form of the
potential to fit baryonic properties and evaluated the appropriate
values of ${\Lambda}$ and ${\Lambda}^{\prime}$. We have borrowed
the same values ($\Lambda \sim 100~MeV$ and $\Lambda^\prime \sim
350~MeV$) in our present calculation.

Ray et al. (2000b) calculated finite temperature effects on
strange matter and strange stars, using the original one parameter
Richardson potential. There, the temperature dependence of gluon
mass was also not considered. We have incorporated this effect
following Alexanian and Nair (1995), who obtained a magnetic
screening for the quark-gluon plasma at high density after
performing self-consistence summation of one loop self-energy. It
has the form $M\approx(2.384)\frac{Cg^2T}{4\pi}$ where $C~=~N$
for SU(N) gauge theory and g is the strong coupling constant
$\frac{g^2}{4\pi}~={\alpha_0}$. Adding this term to the usual
density dependent one, the inverse screening length becomes:

\begin{equation}
(D^{-1})^2 = \frac{ 2 \alpha_0}{\pi} \sum_{i=u,d,s,}k^f_i
\sqrt{(k^f_i)^2 + m_i^2} ~+ 7.152 ~\alpha_0 ~T \label{eq:gmt}
\end{equation}
where we have used our standard notations as before. We have
incorporated this form of the gluon mass (inverse screening
length) for calculations of our new sets of SS EOSs.

\section{Results}\label{sec:result}

\begin{figure}
\resizebox{\hsize}{!}{\includegraphics{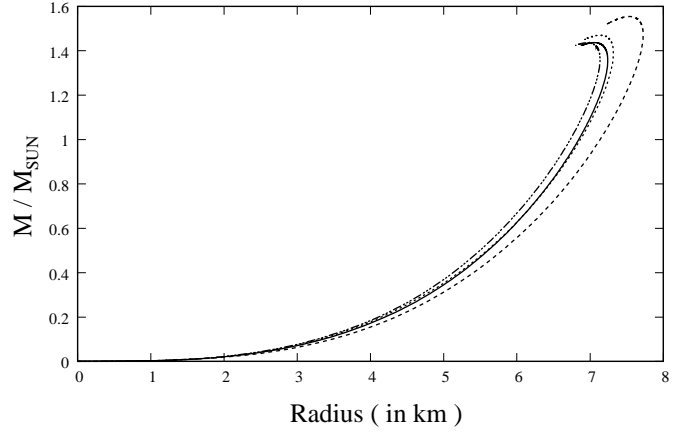}}
\caption{Mass - Radius relations for different parameter sets.
Dashed curve is for EOS C, dotted curve is for EOS B, solid curve
is for EOS SS1 and the dot - dashed curve is for EOS F. }
\label{mr}
\end{figure}

We check our new potential by calculating strange star properties
at zero temperatures. We obtain different EOSs for different set
of parameters tabulated in Table~\ref{tab:alleos}. For
comparison, we also present one EOS SS1 obtained by Dey et al.
(1998) using the unmodified potential in Table~\ref{eos}. In
Figure~\ref{mr}, we compared the M-R plots for three of our new
EOSs with SS1. The compactness (M/R ratio) of the stars has
increased for the new EOSs than compared to SS1.

From Tables~\ref{tab:alleos} and \ref{eos}, the nature of EOSs
and M - R relations do not change appreciably when using the
modified potential if we adjust other model parameters suitably
(see Figure \ref{mr}).

In our new EOSs, we have kept the value of the parameter $N$ to be
the same (except EOS C, where we changed $N$ to see the effects).
However, the value of the parameter ${\alpha_0}$ for our new EOSs
is significantly different from SS1. Here, we chose a larger
value because, as the confinement parameter has now increased from
the previous value of 100 $MeV$ to 350 $MeV$, ${\alpha_0}$ should
be increased to incorporate an earlier onset of deconfinement.
The present value of $\alpha_0$ is 0.55-0.65 compared to 0.20 of
Dey et al., (1998). Such a value is in agreement with theoretical
results (\cite{denalph}). From Table \ref{tab:alleos}, we infer
that the parameter $N$ has the greatest effect on the star
structure and $M_Q$ the least.

\begin{figure}
\resizebox{\hsize}{!}{\includegraphics{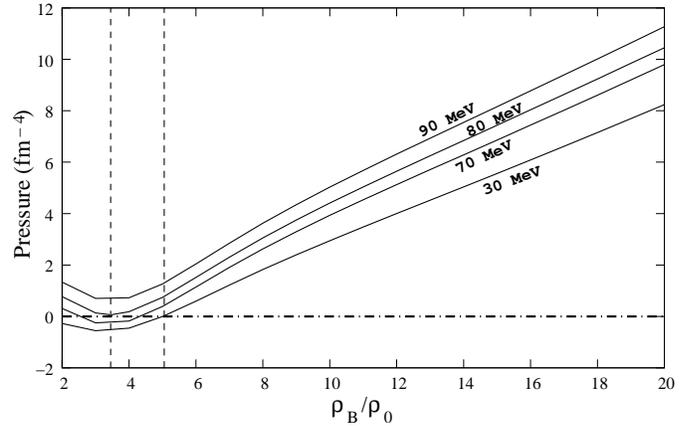}}
\caption{EOSs at different temperatures, all with the same
parameters as EOS F. The first vertical line from the left shows
the location of the zero pressure (and hence stellar surface) for
the EOS at 80 $MeV$, which is around 3.45. The second vertical
line is for an EOS at 30 $MeV$, where the zero of the pressure is
at 5.05.} \label{eostempn}
\end{figure}

\begin{figure}
\resizebox{\hsize}{!}{\includegraphics{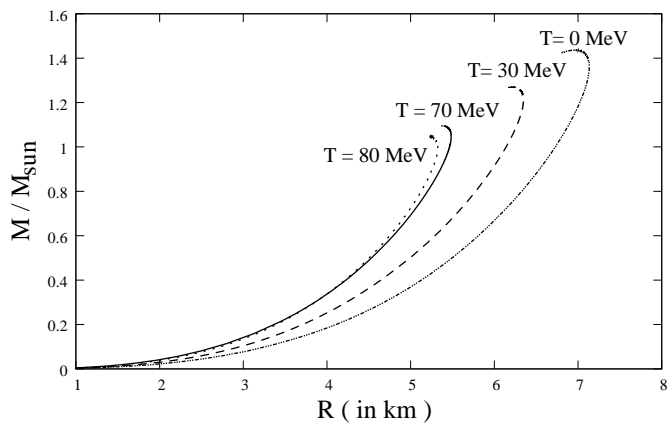}}
\caption{Mass - Radius relations at different temperatures, all
with the same parameters as EOS F.} \label{mrtemp}
\end{figure}

With the parameters used to obtain EOS F, we have calculated
strange star properties at finite temperatures. We have found that
stable stellar structure can be obtained up to a temperature of 80
$MeV$ which is closer to Witten's conjectured temperature of
cosmic separation of phases (100 $MeV$). Our results are shown in
Figures \ref{eostempn} and \ref{mrtemp}. The figures imply that in
our strange star, the density ($\rho_B$) cannot be less than
$4~\rho_0$ because negative pressure is physically not feasible.
This means that the strange star has a sharp surface at a density
$\rho_B \sim 4~\rho_0$ where the pressure goes to zero. This
property of the presence of sharp surfaces in strange stars has
also been seen for other models, like the bag model, the
chromo-dielectric model, etc.

\begin{figure}
\resizebox{\hsize}{!}{\includegraphics{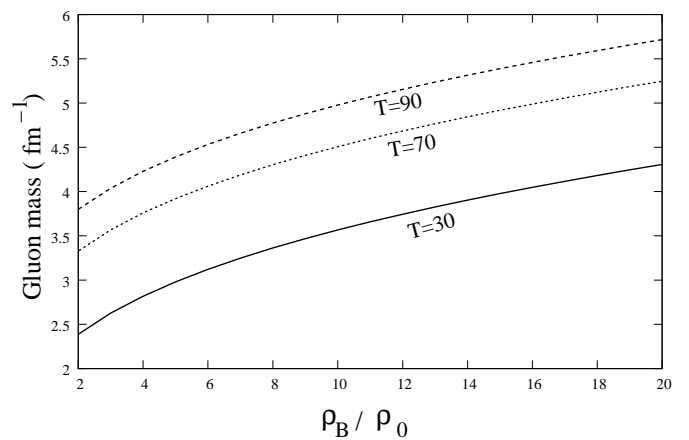}}
\caption{Density and temperature dependence of gluon mass.}
\label{gmasstemp}
\end{figure}

In Figure \ref{gmasstemp}, we have shown both the density and
temperature dependence of gluon mass. It is clear that at a given
temperature, gluon mass increases with density and at a given
density, gluon mass increases with temperature.

\begin{table*}[tb]
\begin{center}
\caption{Data for different EOSs, each having two different values
for ${\Lambda}$ and ${\Lambda}^{\prime}$; ${\Lambda}$ is always
taken as 100 $MeV$. Change in the confinement scale
${\Lambda}^{\prime}$ does not change the minimum energy per
baryon significantly.} \label{tab:alleos}
\begin{tabular}{c c c c c c c c}
\hline \hline
EOS&${\Lambda}^{\prime}$&$M_q$&$N$&${\alpha_0}$&${(E/A)}_{min}$
& $M_{max}$& R \\
Label  & ($MeV$) & ($MeV$) & & & ($MeV$) & ($M_\odot$) &
(km)  \\
&&&&&&&\\ \hline
A& 350&325 &3.0 &.55 &874.0 & 1.532 &7.411  \\

B& 350&325 &3.0 &.65  &907.0&1.470 &7.140 \\

C& 300&325 &2.7 &.55 &877.0  &1.554 &7.530  \\

D & 300&325 &3.0 &.55 &906.0  &1.463 &7.110 \\

E &300&335 &3.0 &.55 &912.0  &1.447 &7.022  \\

F &350&345 &3.0 &.65 &920.3 & 1.436& 6.974  \\
\hline
\end{tabular}
\end{center}
\end{table*}

\begin{table*}[tb]
\begin{center}
\caption{One EOS (SS1) obtained by Dey et al.. (1998) using an
unmodified Richardson Potential with both ${\Lambda}$ and
${\Lambda}^{\prime}$ equal to 100 $MeV$.} \label{eos}
\begin{tabular}{c c c c c c c c }
\hline \hline
EOS &${\Lambda}$&$M_q$&$N$&${\alpha_0}$&${(E/A)}_{min}$&$M_{max}$&$R$\\
Label &($MeV$) & ($MeV$) & & & ($MeV$) & (${M}_ {\odot}$) & (km) \\
\hline
SS1 & 100&310 &3 &.20 & 889.0 & 1.437 &7.06  \\
\hline
\end{tabular}
\end{center}
\end{table*}

\section{Conclusions and summary}\label{sec:conc}
A set of new EOSs for strange matter is presented - both for zero
and very high $T$, using the Richardson potential with the value
of ${\Lambda}^{\prime}~ \simeq~300~{\rm to}~350~MeV$ and the value
of ${\Lambda}~= ~100~MeV$, the two scales for confinement and AF
respectively. This explains both the properties of the deconfined
quark matter (our present work) and the properties of confined 3u
and 3s baryons (Bagchi et al., 2004). Although the confinement
part ${\Lambda}^{\prime}~$ is stronger, leading to a sharp
surface, it is softened by medium effects, developing a screening
length. Inside the star, the AF part is more important. We have
found that for a wide range of parametric variation, the strange
matter EOS gives a minimum of energy per baryon (E/A) that is
lower than ${(E/A)}_{Fe^{56}}=~930.6~MeV$ which ensures that the
system is stable, and unlike neutron star - like structures, our
strange stars are not solely bound by gravitational forces but
also by strong forces. Moreover, considering temperature
dependent screening of the potential, we have found that strange
stars can sustain stable configurations up to a temperature of 80
$MeV$.

The indirect proof of the existence of strange stars is becoming
more prominent because there are many observed features from
compact objects that cannot be explained by the standard neutron
star model, like spectral features and the controversies built
around the star RX J1856.5$-$3754. Another feature, superbursts,
faces difficulties in using the standard carbon burning scenario
in the neutron star crust. The carbon, nitrogen and oxygen mass
fraction ($Z_{CNO}$) must be larger than $Z_{CNO,\odot}$, where
the latter refers to the standard value found in the sun (Cooper
et al. 2005). Sinha et al. (2002) predicted that an alternative
to the carbon burning scenario in a neutron star to explain
superbursts, can be formation of diquark pairs on the strange star
surface and their subsequent breaking, thus giving a continuous
and prolonged emission of energy that can be comparable to the
superbursts. Recently, Page and Cumming (2005) tried to explain
superbursts as carbon burning of matter accumulated on the
electron cloud of a strange star. There are many other features
that can be fitted better with a strange star than a neutron star
model. So, the study of strange matter and strange stars is
important and the formulation of a realistic strange star model
is needed.

\begin{acknowledgements}MB, MD and JD acknowledge DST grant no.
SP/S2/K-03/2001, Govt. of India and the IUCAA, Pune, India, for a
fruitful visit. \end{acknowledgements}

%%%%%%%%%%%%%%%%%%%%%%%%%%%%%%%%%%%%%%%%%%%%%%%%%%%%%%%%%%%%%%%%%%%%%%%%%

%%%                REFERENCES

%%%%%%%%%%%%%%%%%%%%%%%%%%%%%%%%%%%%%%%%%%%%%%%%%%%%%%%%%%%%%%%%%%%%%%%%%

\end{document}